\documentclass[twocolumn, superscriptaddress,amsmath,amssymb,
floatfix,aps,prl,showkeys]{revtex4}
\usepackage{graphics,graphicx}

\usepackage[usenames]{color}
\definecolor{BrickRed}{cmyk}{0,0.89,0.94,0.28}
\definecolor{MidnightBlue}{cmyk}{0.98,0.13,0,0.43}
\definecolor{DarkGreen}{rgb}{0,0.7,0.1}
\newcommand{\comm}[1]{{}}

\begin{document}

\title{The influence of T cell development on pathogen specificity and autoreactivity}

\author{Andrej Ko\v smrlj}
\affiliation{Ragon Institute of MGH, MIT, \& Harvard, Boston, MA 02129, USA}
\affiliation{Department of Physics, Massachusetts Institute of Technology,
Cambridge, MA 02139, USA}

\author{Mehran Kardar}
\affiliation{Department of Physics, Massachusetts Institute of Technology,
Cambridge, MA 02139, USA}

\author{Arup K. Chakraborty}
\email{arupc@mit.edu}
\affiliation{Ragon Institute of MGH, MIT, \& Harvard, Boston, MA 02129, USA}
\affiliation{Departments of Chemical Engineering, Chemistry and
Biological Engineering, Massachusetts Institute of Technology,
Cambridge, MA 02139, USA}


\begin{abstract}
T cells orchestrate adaptive immune responses upon activation. T cell activation requires sufficiently strong binding of T cell receptors on their surface to short peptides derived from foreign proteins bound to protein products of the major histocompatibility (MHC) gene products, which are displayed on the surface of antigen presenting cells. T cells can also interact with peptide-MHC complexes, where the peptide is derived from host (self) proteins. A diverse repertoire of relatively self-tolerant T cell receptors is selected in the thymus. We study a model, computationally and analytically, to describe how thymic selection shapes the repertoire of T cell receptors, such that T cell receptor recognition of pathogenic peptides is both specific and degenerate. We also discuss the escape probability of autoimmune T cells from the thymus.
\end{abstract}

\maketitle
\section{Introduction}
Despite constant exposure to infectious microbial pathogens, higher organisms are rarely sick. This is because the innate immune system is quite successful in clearing pathogens before they can establish an infection. However, the components of the innate immune system respond only to common evolutionary conserved markers expressed by diverse pathogens. Some bacteria and most viruses have evolved strategies to evade or overcome these innate mechanisms of protection.  A second arm of the immune system, adaptive immunity, combats pathogens that escape the innate immune response.  The adaptive immune system is remarkable in that it mounts pathogen-specific responses against a diverse and evolving world of microbes, for which pathogen specificity cannot be pre-programmed.  During an infection, cells of the adaptive immune system that are specific for the pathogen proliferate.  Once the infection is cleared, most of these cells die, but some remain as memory cells which respond rapidly and robustly to re-infection by the same pathogen.  This immunological memory is the basis for vaccination.

T cells play a key role in orchestrating the adaptive immune response.  They combat pathogens that have invaded host cells.  Pathogen-derived proteins in infected host cells are processed into short peptides (p) which can bind to major histocompatibility (MHC) proteins expressed in most host cells.  The resulting pMHC complexes are presented on the surface of host cells as molecular markers of the pathogen. T cells express a protein on their surface called the T-cell receptor (TCR).   Each T-cell receptor (TCR) has a conserved region participating in the signaling functions and a highly variable segment responsible for pathogen recognition. Because the variable regions are generated by stochastic rearrangement of the relevant genes, most T cells express a distinct TCR.  When we say that a given T cell recognizes a particular pMHC complex, we mean that its TCR binds sufficiently strongly to it to enable biochemical reactions inside the T cell that result in activation and proliferation of this particular ``pathogen-specific'' T cell clone.

The diversity of the T cell repertoire enables the immune system to recognize many different pathogenic pMHCs. Peptides presented on MHC class I are typically 8--11 amino acids long~\cite{janewayB}. TCR recognition of pMHC is both degenerate and specific. It is degenerate, because each TCR can recognize several peptides~\cite{unanue84}. It is specific, because most point mutations to the recognized peptide amino acids abrogate recognition~\cite{huseby05, huseby06}. 

Host proteins (e.g., those that are misfolded) are also processed into short peptides, and are presented on the surface of cells in complex with MHC proteins.  The gene rearrangement process ensuring the diversity of TCR may result in generating T cells harmful to the host, because they bind strongly to self pMHCs.  T cells which bind too weakly to MHC are also not useful as they cannot interact with pathogenic peptides. Such T cells, which bind too weakly or too strongly to self pMHC molecules are likely to be eliminated during T cell development in the thymus~\cite{boehmer03, werlen03, hogquist05, siggs06}. Immature T cells (thymocytes) move around the thymus and interact with a diverse set (10$^3$-10$^4$) of self pMHCs presented on thymic cells. Thymocytes expressing TCRs that bind too strongly to any self-pMHC are likely to be deleted (a process called negative selection). However, a thymocyteÕs TCR must also bind sufficiently strongly to at least one self-pMHC to receive a survival signal and emerge from the thymus (a process called positive selection).  The threshold binding strength required for positive selection is weaker than that which is likely to result in negative selection.

Signaling events, gene transcription programs, and cell migration during T cell development in the thymus have been studied extensively~\cite{bousso02, borghans03, boehmer03, werlen03, scherer04, hogquist05, daniels06, siggs06, detours99pnas, detours99tb}. An understanding of how interactions with self-pMHC complexes in the thymus shape the peptide binding properties of selected TCR amino acid sequences such that mature T cells exhibit their special properties is also beginning to emerge. Recently experiments carried out by Huseby et al.~\cite{huseby05, huseby06} provided important clues in this regard.  Experiments were carried out to contrast T cells that developed in a normal mouse with a diversity of self pMHC molecules in the thymus and those that developed in a mouse that was engineered to express only one type of self pMHC in the thymus. For T cells that developed in a normal mouse, pathogen recognition was found to be very sensitive to most point mutations of recognized pathogenic peptides. In contrast, T cells that developed in the engineered mouse were found to be much more cross-reactive.

\begin{figure*}[t]
\begin{center}
\includegraphics[scale=.5]{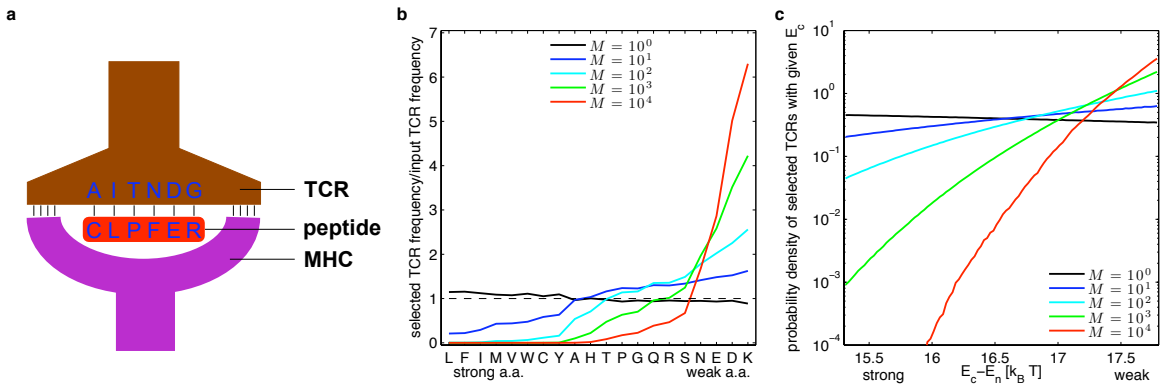}
\caption{
Effects of thymic selection on the characteristics of selected TCRs. a) Schematic representation of the interface between TCR and pMHCs. The region of the TCR contacting the peptide is highly variable and is modeled by strings of amino acids of length $N$. The peptide is also treated similarly. The binding free energy between the TCR and the entire pMHC is computed as described in the text. b) Amino acid composition of TCR selected against $M$ types of self peptides in the thymus. The ordinate is the ratio of the frequency of occurrence of an amino acid in the peptide contact residues of selected TCRs and the pre-selection frequency. TCRs selected against many types of self-peptides in the thymus have peptide contact residues that are enriched in amino acids that interact weakly with other amino acids. Amino acids on the abscissa are ordered according to their largest interaction strength with other amino acids in the potential matrix,  $J$. c) Probability density distribution of $E_c$ values (strength of TCR binding to MHC) of TCRs selected against $M$ types of self peptides. TCRs selected against many types of self peptides are more likely to bind weakly to MHC.  The parameter values are: $N=5$, $E_p - E_n =2.5 k_B T$, $E_{c,\rm{max}} = E_p - N \overline{J}$, $E_{c,\rm{min}} = E_n - N \overline{J}$. We have used the Miyazawa-Jernigan matrix,  $J$~\cite{miyazawa96} and amino acid frequencies $f_a$ from the human proteome~\cite{kosmrlj08,flicek08} (using the mouse proteome does not change the qualitative results~\cite{kosmrlj08}).
} 
\label{fig1}
\end{center}
\end{figure*}

To address these issues, we previously studied a simple model, where TCRs and pMHCs were represented by strings of amino acids~\cite{chao05} (Fig.~\ref{fig1}a), using numerical~\cite{kosmrlj08} and analytical~\cite{kosmrlj09} methods.  Our results provided a statistical perspective on the origin of how T cells recognize foreign pathogens in a specific, yet degenerate, manner.  They also provided a conceptual framework for diverse experimental data~\cite{kosmrlj10, eisen10}.  In this paper, we extend the model, and study new phenomena that include:  1) How TCR-MHC interactions differ upon development against different numbers of peptides in the thymus, and how this influences T cell cross-reactivity?  2) What are the sequence characteristics of pathogenic peptides recognized by T cells?  3) How does stochastic escape from negative selection in a normal thymus influence T cell specificity for pathogenic peptides?  4) How does the frequency of autoimmune T cells change upon modulating the number of peptides encountered during T cell development? 

\section{Model}
To describe the interactions between TCRs and pMHCs, we model them as strings of amino acids. These strings indicate the amino acids on the interface between TCRs and pMHCs.  In the simplest incarnations of the model, it is assumed that each site on a TCR interacts only with a corresponding site on a pMHC (Fig.~\ref{fig1}a). The binding interface of a TCR is composed of a more conserved region that is in contact with the MHC molecule and a highly variable region that makes the majority of contacts with the peptide. Therefore, we explicitly consider only amino acids of the latter part of the TCR, but not the former. Similarly, there are many possible peptides that can bind to MHC, and their sequences are considered explicitly. Prior to our work~\cite{kosmrlj08, kosmrlj09, kosmrlj10}, TCR-pMHC interactions have been represented using string models~\cite{detours99pnas, detours99tb, chao05}, but these studies did not have an explicit treatment of amino acids or consider the mechanistic issues we did (including connections to human disease~\cite{kosmrlj10}).

To assess the effects of thymic development on pathogen recognition characteristics, we evaluate the free energy of interaction between TCR-pMHC pairs. The interaction free energy is composed of two parts: a TCR interaction with MHC and a TCR interaction with the peptide. The former is given a value $E_c$, which may be varied to describe different TCRs and MHCs. The latter is obtained by aligning the TCR and pMHC amino acids that are treated explicitly and adding the pairwise interactions between corresponding pairs. For a given TCR-pMHC pair, this gives
\begin{equation}
E_\mathrm{int}\left(E_c, \vec{t}, \vec{s}\, \right) = E_c + \sum_{i=1}^N J\left(t_i,s_i \right),
\label{eq1}
\end{equation}
where $J\left(t_i,s_i\right)$ is the contribution from the $i$th amino acid of the TCR ($t_i$) and the peptide ($s_i$) and $N\sim5$ is the length of the variable TCR-peptide region. The matrix $J$ encodes the interaction energies between specific pairs of amino acids. For numerical purposes we use the Miyazawa-Jernigan matrix~\cite{miyazawa96} that was developed in the context of protein folding, but as will be described later the qualitative results do not depend on the form of $J$.

To model thymic selection, we start by randomly generating a set of $M$ self peptides, where amino acids are picked with frequencies corresponding to the human proteome~\cite{kosmrlj08, flicek08} (using the mouse proteome does not change the qualitative results~\cite{kosmrlj08}). Then we randomly generate TCR sequences with the same amino acid frequencies. To mimic thymic selection, TCR sequences that bind to any self-pMHC too strongly ($E_\mathrm{int} < E_n$) are deleted (negative selection). However, a TCR must also bind sufficiently strongly ($E_\mathrm{int} < E_p$) to at least one self-pMHC to receive survival signals and emerge from the thymus (positive selection). Recent experiments show that the difference between the thresholds for positive and negative selection is relatively small (a few $k_B T$)~\cite{daniels06}. The threshold for negative selection ($E_n$) is quite sharp, while the threshold for positive selection ($E_p$) is soft. Replacing soft thresholds with perfectly sharp thresholds at $E_n$ and $E_p$ does not change the qualitative behavior of the selected T cell repertoire (see below and ref.~\cite{kosmrlj08}). However, we do carry out calculations with soft thresholds as well to study the escape of potentially autoimmune T cells, and their pathogen recognition characteristics.

To completely specify the interaction free energy between a TCR and pMHC, we need to specify the value of $E_c$. In previous studies~\cite{kosmrlj08, kosmrlj09} we fixed the value of $E_c$ for all TCRs at some moderate value, because too strong binding to MHC (large $|E_c|$) would result in negative selection with any peptide, and too weak a binding to MHC (small $|E_c|$) would result in TCR not being positively selected.  Each human can have up to 12 different MHC types.  A TCR that binds strongly to more than one MHC type is likely to be eliminated during negative selection.  Therefore, we consider TCRs that are restricted by a particular MHC type.   We expect that variations in $E_c$ for selected TCRs are small. A rough estimate on the bounds can be obtained from the condition that the average interaction free energy between TCR and pMHC for selected TCRs is between the thresholds for positive and negative selection
\begin{equation}
E_n < E_c + N \overline{J} < E_p,
\label{eq2}
\end{equation}
where $\overline{J}$ is the average value of interaction between amino acids. The upper (lower) bound $E_{c,\mathrm{max}} = E_p - N \overline{J}$ ($E_{c,\mathrm{min}} = E_n - N \overline{J}$) ensures that, on average, interactions result in positive selection and not negative selection. Since for selection it is enough that a TCR sequence is positively selected by one of many self peptides and avoid being selected by encountered self peptides, the actual bounds for $E_c$ might be different, but we expect that the range of $E_c$ values is still small; viz., $E_{c,\mathrm{max}} - E_{c,\mathrm{min}} \propto E_p - E_n$. To every TCR sequence we assign a random value of $E_c$ chosen uniformly from the interval $\left(E_{c,\mathrm{min}}, E_{c,\mathrm{max}} \right)$, and subject it to the selection processes. Note that TCRs with interactions with MHCs that are too weak are unlikely to be oriented on MHCs properly and hence will be unable to interact with the peptide.  Thus, one cannot tune $E_c$ to very low values to escape negative selection.

\section{Results}
\subsection{TCRs selected against many self peptides are enriched with weakly interacting amino acids and bind more weakly to MHCs}

First, we study how thymic selection shapes TCR sequences and TCR interactions with MHC. A million randomly generated TCR peptide contact residues with randomly assigned $E_c$ values are generated and selected against $M$ randomly generated self peptides according to the thymic selection rules described in the previous section. For the set of selected TCRs, we assess their amino acid composition and their interactions with MHC ($E_c$ values). The whole process is repeated thousand times to obtain proper statistics.  The peptide contact residues of TCR sequences selected against many self peptides are statistically enriched with weakly interacting amino acids (Fig.~\ref{fig1}b), and TCRs with weaker binding to MHC (within the allowed range) are more likely to get selected (Fig.~\ref{fig1}c). This is because negative selection imposes a strong constraint. When selected against many self peptides, TCR sequences with peptide contact residues containing strongly interacting amino acids (e.g., hydrophobic amino acids or those with flexible side chains) or TCRs that bind strongly to MHC are more likely to bind strongly with at least one self-pMHC and thus be negatively selected. This qualitative result agrees with experiment~\cite{huseby06, kosmrlj08} and is independent of details of the interaction potential $J$ or the sharpness of the thresholds for positive and negative selection as will be shown next.

The selection of a given TCR sequence $\vec t\, $ is determined by the strongest interaction with all self peptides, and a TCR is selected when
\begin{equation}
E_n < \min_{\vec s \in M} \left\{ E_\mathrm{int} \left( E_c, \vec t, \vec s \, \right) \right\}< E_p.
\label{eq3}
\end{equation}
In Ref.~\cite{kosmrlj09} we showed that by using the Extreme Value Distribution one finds that the strongest interaction energy with $M$ random self peptides is sharply peaked around 
\begin{equation}
E_0 \left( E_c, \vec t\, \right) = E_c + \sum_{i=1}^N \mathcal{E}(t_i) - \sqrt{(2 \ln M) \sum_{i=1}^N \mathcal{V} (t_i)},
\label{eq4}
\end{equation}
where $\mathcal{E}(a) = \left[ J(t_i, a) \right]_a$ and $\nu (a) = \left[ J(t_i, a)^2 \right]_a - \left[ J(t_i, a) \right]_a^2$ are the average and the variance of the interaction free energy of amino acid a with all others. We have denoted the average over self amino acid frequencies by $\left[ G(a) \right] \equiv \sum_{a=1}^{20} f_a G(a)$. From this equation and the selection condition (\ref{eq3}) we see that, as the number of self peptides, $M$, increases, the chance of negative selection does too. To counterbalance this pressure for large $M$, TCRs are enriched with weakly interacting amino acids in their peptide contact residues and TCRs that interact weakly with MHC (small $E_c$ value) (see Fig.~\ref{fig1}). A similar effect can be obtained with amino acids with smaller variance of interactions, but this effect is less pronounced because of the square root.  Even if it were in effect, it would pick out TCRs with smaller variance, which for the case where the means were the same, also imply selecting the more weakly binding amino acids.  These results are independent of the form of the statistical potential between contacting amino acids. Different potentials only change the identities of weak and strong amino acids.

The probabilities with which amino acids are chosen for the selected TCRs in the T cell repertoire depend on the conditions (e.g. the number of peptides present in the thymus). This dependency can be formalized by using statistical mechanical methods that apply in the limit of very long peptides and remarkably the results seem to be accurate even for short peptides~\cite{kosmrlj09}. The thymic selection condition 
\begin{equation}
E_n < E_0 \left( E_c, \vec t\,  \right) < E_p
\label{eq5}
\end{equation}
can be interpreted as a micro-canonical ensemble of sequences $\vec t\,$, which are acceptable if the value of the Hamiltonian, $E_0 \left( E_c, \vec t \, \right)$, falls on the interval $(E_n, E_p)$. In the limit of long peptides canonical and micro-canonical ensembles are equivalent. Thus the probability for TCR selection is governed by the Boltzmann weight
\begin{equation}
p\left(E_c, \vec t\, \right) \propto \left( \prod_{i=1}^N f_{t_i} \right) \rho(E_c) \exp\!\bigg[-  \beta E_0 \left( E_c, \vec t\, \right)\bigg].
\label{eq6}
\end{equation}
Here $f_a$ and $\rho(E_c)$ are the natural frequencies of the different amino acids and the distribution of $E_c$ values prior to selection, whereas the effect of thymic selection is captured in the parameter $\beta$, which is determined by the condition that the average free energy falls in the interval $(E_n, E_p)$. The complication, presented by the square root term in Eq.~(\ref{eq4}), for determining parameter $\beta$ is easily dealt with by Hamiltonian minimization~\cite{kardar83} and introducing an effective Hamiltonian,
\begin{equation}
H_0 \left( E_c, \vec t\, \right) = E_c + \sum_{i=1}^N \big[ \mathcal{E} (t_i) - \gamma \mathcal{V} (t_i) \big] - \ln M/(2 \gamma).
\label{eq7}
\end{equation}
This corresponds to Boltzmann weights
\begin{equation}
\begin{array}{r l}
p\left(E_c, \vec t\, \right) \propto & \rho(E_c) \exp\left[- \beta E_c\right] \cr
 & \times \prod_{i=1}^N \left\{ f_{t_i} \exp\left[ - \beta \left(\mathcal{E}(t_i) - \gamma \mathcal{V} (t_i) \right) \right]\right\} \cr
\end{array}
\label{eq8}
\end{equation}
for which thermodynamic quantities are easily computed. $\gamma (\beta) = \sqrt{\ln M / \left( 2 N \langle \mathcal{V} \rangle \right)}$ is determined by minimizing the effective Hamiltonian $H_0 \left( E_c, \vec t\,  \right)$ with respect to $\gamma$, which ensures that the average free energies $\langle E_0 \left( E_c, \vec t\,  \right) \rangle$ and $\langle H_0 \left( E_c, \vec t\,  \right) \rangle$ are the same; the averaging is done over all TCR sequences $\vec t$ weighted with Boltzmann weights (\ref{eq6}) and (\ref{eq8}).

\begin{figure*}[t]
\includegraphics[scale=.5]{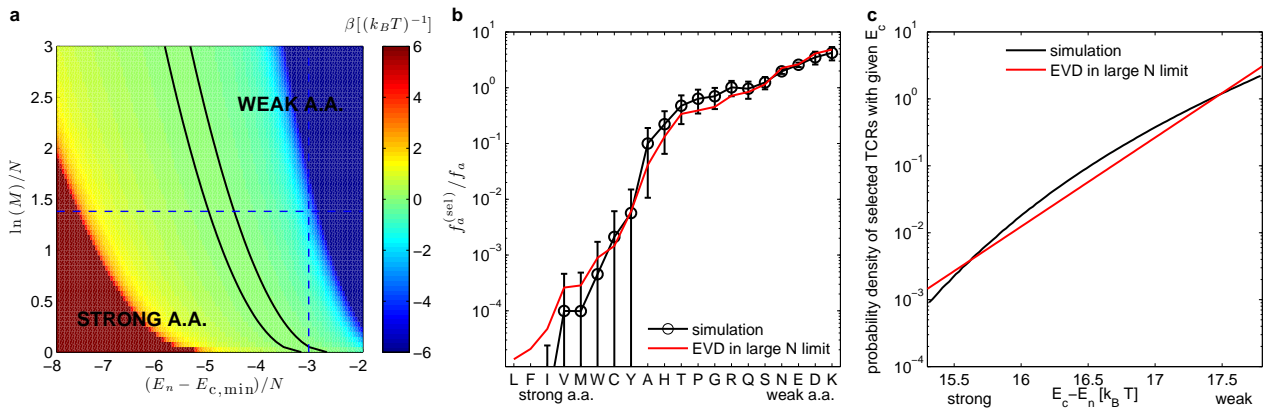}
\caption{
Analytical results for characteristics of selected TCRs. (a) Representation of the dependency of the parameter $\beta$ (a measure of amino acid composition of selected TCRs and a measure of selected TCR binding strengths to MHCs, see text), on the number of self-peptides ($\ln M/N$) and the threshold for negative selection $E_n$ with $(E_p -E_n)/N =(E_{c,\rm{max}} -E_{c,\rm{min}})/N = 0.5 k_B T$. The region between the black lines corresponds to $\beta=0$, to the right (left) of which negative (positive) selection is dominant, and weak (strong) amino acids are selected. The blue dashed lines indicate the relevant parameter values for thymic selection in mouse. (b) Amino-acid composition of selected TCR sequences, ordered in increasing frequency along the abscissa. (c) Probability density distribution of $E_c$ values (strength of TCR binding to MHC) of selected TCRs. (b-c) The data points in black are obtained numerically with the parameters relevant to mouse (see text). The error bars reflect the sample size used to generate the histograms and differences for different realizations of $M$ self-peptides. The red lines are the result of the EVD analysis in the large $N$ limit (from Eqs.~(\ref{eq9}) and (\ref{eq10})), and the agreement is quite good. In both cases we have used the Miyazawa-Jernigan matrix, $J$~\cite{miyazawa96} and amino acid frequencies $f_a$ from the human proteome~\cite{kosmrlj08, flicek08}. } 
\label{fig2}
\end{figure*}

$\beta$ is determined by constraining the average free energy to the range $(E_n, E_p)$, while maximizing entropy. Given the bounded set of free energies, the parameter $\beta$ can be either negative or positive. The values for $E_0 \left( E_c , \vec t\, \right)$ span a range from $E_\mathrm{min}$ to $E_\mathrm{max}$, and a corresponding number of states $\Omega(E_0)$ is bell-shaped between these extremes with a maximum at some $E_\mathrm{mid}$. If $E_\mathrm{mid} > E_p$, we must set $\beta$ such that $\langle E_0 \left( E_c, \vec t\,  \right) \rangle=E_p$. In this case, $\beta>0$, positive selection is dominant and selected TCRs contain peptide contact residues with stronger amino acids and TCRs that interact strongly with MHCs. If $E_\mathrm{mid} < E_n$, we must set $\beta$ such that $\langle E_0 \left( E_c, \vec t\,  \right) \rangle=E_n$; now $\beta<0$, negative selection is dominant, and TCRs with peptide contact residues with weaker amino acids and TCRs that interact weakly with MHCs are selected. For $E_n <E_\mathrm{mid} < E_p$, we must set $\beta=0$ and there is no modification due to selection. The resulting phase diagram of parameter $\beta$ is shown in Figure~\ref{fig2}a.

For the relevant parameters in mouse (i.e. $N=5$, $E_p - E_n = 2.5 k_B T$, $E_{c,\mathrm{max}} = E_p - N \overline J$, $E_{c,\mathrm{min}} = E_n - N \overline J$, and $M=10^3$), we find $\beta=-3.06 (k_B T)^{-1} $  and $\gamma=0.94 (k_B T)^{-1}$, negative selection is dominant and weaker amino acids are selected. This result is consistent with experiments~\cite{kosmrlj08}. With these parameters we can calculate the amino acid frequencies of selected TCRs as
\begin{equation}
f_a^{(\mathrm{sel})} = \frac{f_a \exp\big[ - \beta \big( \mathcal{E} (a) - \gamma \mathcal{V} (a) \big) \big]}{\sum_{b=1}^{20} f_b \exp\big[ - \beta \big( \mathcal{E} (b) - \gamma \mathcal{V} (b) \big) \big]}
\label{eq9}
\end{equation}
and the distribution of selected TCRs interactions with MHCs as
\begin{equation}
\rho^{(\mathrm{sel})} (E_c) = \frac{\rho(E_c) \exp\left[ - \beta E_c \right]}{\int_{E_{c,\mathrm{min}}}^{E_{c,\mathrm{max}}} \rho(E) \exp[-\beta E] dE},
\label{eq10}
\end{equation}
where $\rho(E_c)$ is the distribution of TCR interactions with MHCs before selection, which is taken to be a uniform distribution in our simulations. We find (Figs.~\ref{fig2}b and~\ref{fig2}c) that the analytical results above agree very well with the numerical results of simulations for $N=5$ and the parameters presented above.

\subsection{Selection against many self peptides is required for pathogen-specific T cells}

\begin{figure*}[t]
\begin{center}
\includegraphics[scale=.5]{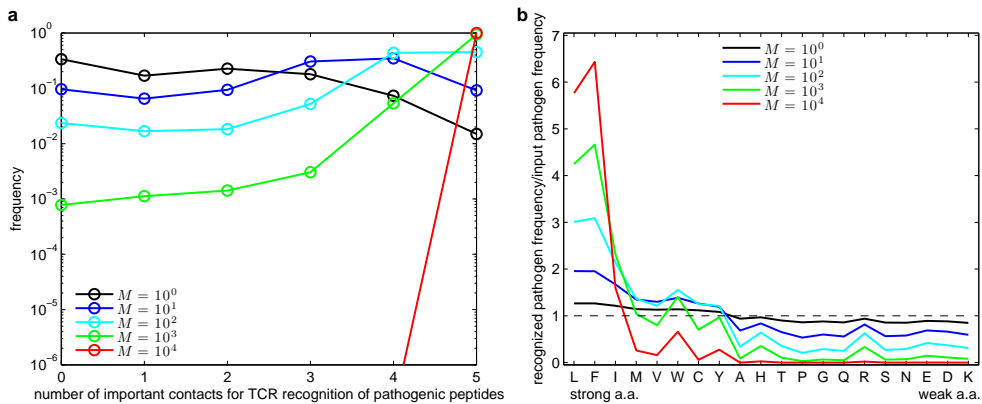}
\caption{
Characteristics of TCR recognition of pathogenic peptides. a) Histogram of the number of important contacts (defined in text) with which T cells recognize pathogenic peptides. T cells selected against many self peptides recognize pathogenic peptides via many important contacts and are thus specific. In contrast T cells selected against few types of self peptides recognize pathogenic peptides with only a few important contacts and are thus cross-reactive. b) Amino acid composition of pathogenic peptides that are recognized by at least one of the selected TCRs. TCRs selected against many types of self peptides recognize pathogenic peptides that are enriched with strongly interacting amino acids. In contrast TCRs selected against few types of self peptides may also recognize pathogenic peptides that contain weakly or moderately interacting amino acids. We used the same set of parameters as in Figure~\ref{fig1} to obtain these results.} 
\label{fig3}
\end{center}
\end{figure*}

How does such a T cell repertoire lead to specific recognition of pathogenic peptide? To study the specificity of mature T cells for pathogenic peptide recognition, we challenge selected TCR sequences with a collection of many randomly generated pathogenic peptides where amino acid frequencies correspond to {\it L. monocytogenes}~\cite{kosmrlj08, moszer95}, a pathogen that infects humans. TCR recognition of pathogenic peptide occurs if TCR-pMHC binding is sufficiently strong ($E_\mathrm{int} < E_r$), where the recognition threshold in mouse experiments is such that $E_r \sim E_n$~\cite{naeher07}. For each TCR that recognizes a particular pathogenic sequence, the specificity of recognition was tested. Each site on the peptide was mutated to all other 19 possibilities, and recognition of the reactive TCRs was again assessed. If more than half the mutations at a particular site abrogated recognition by the same TCR, this site was labeled an important contact. For each strongly bound TCR-pMHC pair, the number of important contacts was determined. After summing over all selected TCRs and pathogenic peptides, we obtained a histogram of the number of important contacts (Fig.~\ref{fig3}a). The higher the number of important contacts the more specific is the TCR recognition of pathogenic peptide. Small numbers of important contacts correspond to cross-reactive TCRs that are able to recognize many pathogenic peptide mutants. The obtained result is qualitatively the same as the one obtained in previous studies~\cite{kosmrlj08, kosmrlj09}, where the binding free energy of TCRs with MHC was fixed.

In agreement with experiments~\cite{huseby05, huseby06} we find in our model that TCRs selected against many different self peptides are very specific, while TCRs selected against only one self peptide are more cross-reactive (Fig.~\ref{fig3}a). Based on the amino acid composition of selected TCRs (see previous section), we can provide a mechanistic explanation for the specificity/degeneracy of pathogen recognition. Because TCR peptide contact residues are enriched with weakly interacting amino acids and TCRs are more likely to react moderately to MHC, they can interact sufficiently strongly for recognition to occur only with pathogenic peptides that are statistically enriched in amino acids that are the stronger binding complements of the peptide contact residues of the TCR (Fig.~\ref{fig3}b). Such TCR-peptide interactions generate weak to moderate interactions which sum up to provide sufficient binding strength for recognition; each interaction contributes a significant percentage of the total binding affinity. If there is a mutation to a peptide amino acid of a recognized peptide, it is likely to weaken the interaction it participates in as recognized peptides are statistically enriched in amino acids that interact strongly with the TCR's peptide contact residues. Weakening an interaction that contributes a significant percentage of the binding free energy is likely to abrogate recognition because the recognition threshold is sharply defined~\cite{daniels06}.

This statistical view of TCR specificity for antigen may describe the initial step of binding, which may then allow modest conformational adjustments~\cite{eisen10}. This mechanism also suggests an explanation for why TCR recognition of pathogenic peptides can be degenerate. There are many combinatorial ways of distributing strongly interacting amino acids along the peptide, which lead to sufficiently strong binding with TCR for recognition.

In agreement with experiments~\cite{huseby06} for T cells that develop in mice with many peptides in the thymus, sufficiently strong binding for recognition is achieved via many moderate bonds and each of these bonds is important for recognition. In contrast TCR sequences selected against only one type of self-peptide have a higher chance of containing strongly interacting amino acids and have a higher chance to bind more strongly to the MHC (Fig.~\ref{fig1}). Such TCRs can recognize a lot more pathogenic peptides including the ones that contain weakly or moderately interacting amino acids (Fig.~\ref{fig3}b). In many cases mutating such amino acids on the peptide does not prevent recognition of the same TCR because a small number of strong contacts dominate recognition (Fig.~\ref{fig3}a and experiments~\cite{huseby06}), and unless these specific ones are disrupted by mutations to the peptide, recognition is not abrogated. Accordingly TCR recognition of pathogenic peptides is more cross-reactive. 

It may also happen that the binding interaction between a TCR and pathogenic peptide-MHC is sufficiently strong that a single mutation of peptide amino acids cannot prevent recognition, which results in $0$ important contacts. This may happen because of the stronger binding of TCRs to MHC and because of the higher chance of TCRs having strongly interacting peptide contact residues. When selected against fewer types of self-peptides, TCRs that bind strongly to MHC can escape (Fig.~\ref{fig1}c). Thus in this case the escape of TCRs that bind strongly or moderately to more than one MHC type (or MHC with mutations) might also be possible, leading to more cross-reactivity to MHC types (or substitutions of MHC amino acids~\cite{huseby05}).

\subsection{Characteristics of foreign peptides recognized by T cells}

Once T cells complete thymic selection, a set $\mathcal{T}$ of TCRs ($K$ in number) is released in the blood stream, where they try to identify infected cells. A T cell recognizes infected cells when its TCR binds sufficiently strongly ($E_\mathrm{int} < E_r$) to foreign peptide-MHC.  Experimental evidence~\cite{naeher07} suggests that the negative selection threshold in the thymus is the same as the recognition threshold in the periphery, i.e. $E_r \sim E_n$. This means that a foreign peptide of sequence $\vec s\,$  is recognized if \emph{its strongest interaction} with the set of TCRs exceeds the threshold for recognition, i.e. 
\begin{equation}
\min_{\left\{E_c,\vec t\, \right\} \in \mathcal{T}} \left\{ E_\mathrm{int} \left(E_c, \vec t, \vec s\,\right) \right\} < E_n.
\label{eq11}
\end{equation}

Eq.~(\ref{eq11}) casts the recognition of foreign peptides as another extreme value problem. To calculate the probability $P_\mathrm{rec}\left( \vec s\, \right) $ that a foreign peptide sequence $\vec s\, $  is recognized by T cells and to calculate the amino acid composition of recognized foreign peptides in the limit of long peptide sequences, we use the same procedure that was used in Ref.~\cite{kosmrlj09} (briefly discussed in previous section) to calculate the properties of selected TCRs. Therefore, here we just briefly summarize the necessary steps. 

Let us indicate by $\rho^*\left(x | \vec s\, \right)$  the probability density function (PDF) of the interaction free energy between the foreign peptide $\vec s\, $ and a random TCR that is selected in the thymus. Then the probability that foreign peptide $\vec s\, $ is recognized is obtained by integrating the extreme value distribution (EVD) $\Pi^*\left( x | \vec s\, \right)$ over the allowed range:
\begin{eqnarray}
P_\mathrm{rec} \left(\vec s\, \right)& = &  \int_{-\infty}^{E_n} \Pi^*\! \left( x | \vec s\, \right) dx, \quad{\rm with}\nonumber \\
\Pi^* \! \left(x|\vec s \, \right) & = & K \rho^* \left( x | \vec s \, \right) \big[1 - P^*\left(E < x | \vec s \,\right) \big]^{K-1}.
\label{eq12}
\end{eqnarray}
where $P^*$ is the cumulative probability of the PDF $\rho^*$. If we model the set $\mathcal{T}$ of selected T cells as $K$ strings in which each amino acid is chosen independently with frequencies $f_a^{(\mathrm{sel})}$ (i.e. we ignore the correlations among different positions on the string), then in the limit of long peptide sequences (large $N$) we can approximate the PDF $\rho^*\left( x | \vec s\, \right)$  with a Gaussian with mean $E_\mathrm{av}^*(\vec s\,) = \langle E_c \rangle + \sum_{i=1}^N \mathcal{E}^*(s_i)$ and variance $V^*\left( \vec s\, \right) = \langle E_c^2 \rangle_c + \sum_{i=1}^N \mathcal{V}^* (s_i)$. The mean $\mathcal{E}^* (s_i)$ and the variance $\mathcal{V}^* (s_i)$ of the amino acid interaction free energies are obtained as in the previous section by appropriately replacing $f_a$ with $f_a^{(\rm{sel})}$. The mean $\langle E_c \rangle$ and the variance $\langle E_c^2 \rangle_c = \langle E_c^2 \rangle - \langle E_c \rangle^2$ of selected TCR interactions with MHCs are defined as
$\langle X \rangle =
\frac{\int_{E_{c,\rm{min}}}^{E_{c,\rm{max}}} X \rho\left(E_c\right) \exp\left[ -\beta E_c \right] dE_c}
{\int_{E_{c,\rm{min}}}^{E_{c,\rm{max}}}  \rho\left(E_c\right) \exp\left[ -\beta E_c \right] dE_c}$,
where $\rho\left( E_c \right)$ is distribution of $E_c$ values before selection. In the limit of large number of T cells ($K \gg 1$) the extreme value distribution $\Pi^*\left( x | \vec s\, \right)$ is sharply peaked around 
\begin{equation}
\begin{array}{rl}
E_0^* \left( \vec s\, \right) = & \langle E_c \rangle + \sum_{i=1}^N \mathcal{E}^* (s_i) \cr
 & - \sqrt{\left( 2 \ln K \right) \left[ \langle E_c^2 \rangle_c + \sum_{i=1}^N \mathcal{V}^* (s_i) \right]}, \cr
 \end{array}
\label{eq13}
\end{equation}
and in the large $N$ limit the condition for recognition of foreign peptides becomes
\begin{equation}
E_0^*\left( \vec s \right) < E_n .
\label{eq14}
\end{equation}
The probability for a sequence $\vec s\, $ to be recognized is governed by the Boltzmann weight 
$p\left( \vec s\, \right) \propto \left( \prod_{i=1}^N \tilde{f}_{s_i} \right) \exp\left[-\beta^* E_0^* \left( \vec s\, \right) \right]$, where $\left\{ \tilde{f}_a \right\}$ are natural frequencies of different amino acids in the pathogen proteome, while the effect of TCR recognition is captured in the parameter $\beta^*$. As in the previous section, we introduce a new Hamiltonian 
$H_0^* \left( \vec s\, \right) = \langle E_c \rangle - \gamma^* \langle E_c^2 \rangle_c + \sum_{i=1}^N \left[ \mathcal{E}^*(s_i) - \gamma^* \mathcal{V}^* (s_i) \right] - \ln K / \left( 2 \gamma^* \right)$, and to ensure the same average energies $\langle E_0^* \left( \vec s\, \right) \rangle$ and $\langle H_0^* \left( \vec s\, \right) \rangle$ we set $\gamma^* \left(\beta^* \right) = \sqrt{\ln K / \left( 2 \langle E_c^2 \rangle_c + 2 N \langle  \mathcal{V}^* \rangle  \right)}$. Finally, the value of $\beta^*$ is determined by constraining the average energy $\langle E_0^* \left( \vec s \right) \rangle < E_n$, while maximizing entropy. If $E_\mathrm{mid}^* > E_n$, we must set $\beta^*$ such that $\langle E_0^* \left( \vec s \right) \rangle = E_n$, where $E_\mathrm{mid}^*$ is defined as in the previous section. In this case, $\beta^*>0$, and only foreign peptides with stronger amino acids are recognized. If $E_\mathrm{mid}^* < E_n$, we must set $\beta^*=0$, and there is no modification due to recognition, i.e. every foreign peptide is recognized. Note that unlike for the thymic selection of T cell receptors (the 
parameter $\beta$), the parameter $\beta^*$ cannot be negative, because there is no lower energy bound for recognition in Eq.~(\ref{eq14}). With all parameters determined, we can calculate the amino acid frequencies of recognized foreign peptides
\begin{equation}
\tilde{f}_a^{(\rm{rec})} = \frac{\tilde{f}_a \exp\left[ - \beta^* \big( \mathcal{E}^*(a) - \gamma^* \mathcal{V}^*(a) \big) \right]}
{\sum_{b=1}^{20} \tilde{f}_b \exp\left[ - \beta^* \big( \mathcal{E}^*(b) - \gamma^* \mathcal{V}^*(b) \big) \right]}.
\label{eq15}
\end{equation}

\begin{figure*}[t]
\includegraphics[scale=1]{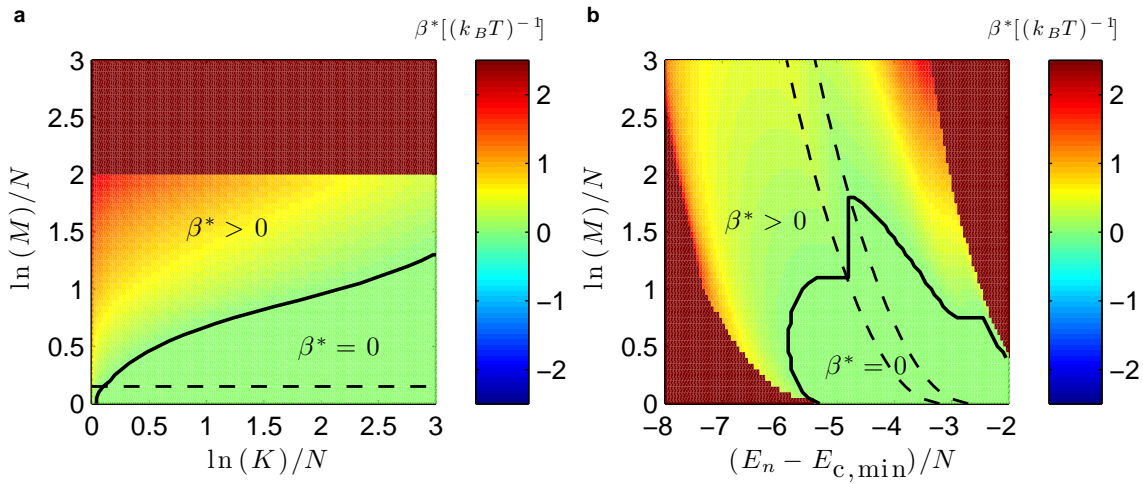}
\caption{
Color representation of the dependence of the parameter $\beta^*$ on the number of selected TCRs ($\ln (K)/N$), the number of self peptides ($\ln (M)/N$) against which TCRs were selected, and threshold for negative selection $E_n/N$.  Parameters are: $(E_p-E_n)/N=0.5 k_B T$ and $E_{c,\rm{min}} - E_{c,\rm{max}} = E_n - E_p$, in the limit of large $N$. Solid black lines separate regions with $\beta^*>0$  and $\beta^*=0$. The regions below the black dashed line in (a) and between the black dashed lines in (b) correspond to $\beta=0$ (every TCR is selected).  In (a) $E_{c,\rm{min}} = E_n - N \overline{J}$ and in (b) $\ln(K)/N=1.5$.} 
\label{fig4}
\end{figure*}

Figure~\ref{fig4} depicts variation of $\beta^*$  as a function of the number of selected TCRs ($\ln (K)/N$), the number of self peptides ($\ln (M)/N$) against which TCRs were selected, and the threshold for negative selection $E_n/N$ with $(E_p-E_n)/N=0.5 k_B T$. The region $\beta^*=0$  is possible only for $K \gtrsim M$  and for parameters where $|\beta|$  is small (cf. Fig.~\ref{fig2}a). That is all foreign peptides are recognized, when there are lots of TCRs or when TCRs are selected against a small number of self peptides, $M$, in the thymus; neither condition is biologically true.

\begin{figure}[h!]
\includegraphics[scale=.4]{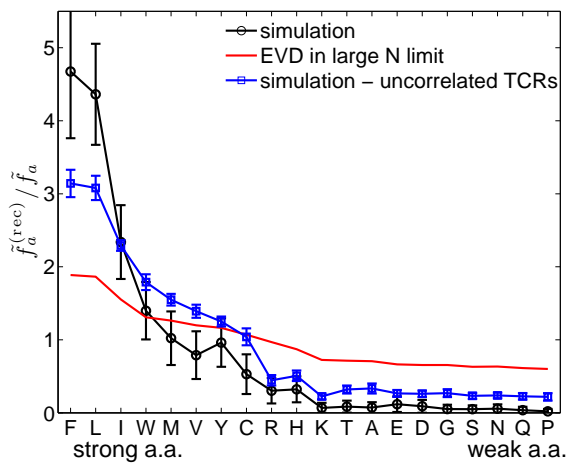}
\caption{
Amino-acid composition of recognized foreign peptides.  The amino acids are ordered in decreasing frequency along the abscissa. The data points in black are obtained numerically with the parameters relevant to TCR selection in the mouse and $K=10^3$ TCRs, which where than challenged with \emph{L. monocytogenes} peptides (see text). The blue data points are for similarly challenged $K=10^3$ uncorrelated TCRs. The error bars reflect the sample size used to generate the histograms and differences for different realizations of $M$ self peptides (black) or $K$ uncorrelated TCRs (blue). The red line is the result of the EVD analysis in the large $N$ limit from Eq.~(\ref{eq15}), where TCR amino acid frequencies were obtained from Eq.~(\ref{eq9}). In all cases we have used the Miyazawa-Jernigan matrix, $J$~\cite{miyazawa96}, amino acid frequencies $f_a$ from the human proteome (using the mouse proteome does not change the qualitative results~\cite{kosmrlj08}), and amino acid frequencies $\tilde{f}_a$ from the \emph{L. monocytogenes proteome}~\cite{kosmrlj08, flicek08}.} 
\label{fig5}
\end{figure}

We also compared the analytical results, which are exact in the limit $N \rightarrow \infty$ with numerical simulations for $N=5$ (Fig.~\ref{fig5}). From the set of parameters that are relevant for thymic selection in the mouse (i.e. $N=5$, $E_p - E_n = 2.5 k_B T$, $E_{c,\rm{max}} = E_p - N \overline{J}$, $E_{c,\rm{min}} = E_n - N \overline{J}$, and $M=10^3$), we generated a pool of $K=10^3$ selected TCRs. Then we randomly generated $10^6$ foreign peptides with amino acid frequencies $\tilde{f}_a$, which were representative of \emph{L. monocytogenes}~\cite{kosmrlj08, moszer95}, and checked the amino acid composition of foreign peptides that were recognized by at least one TCR. We find that selected TCRs recognize only foreign peptides that are enriched with strongly interacting amino acids (Fig.~\ref{fig5}). Increasing experimental evidence indicates that this may be true~\cite{eisen10}. The physical reason for this was discussed in the previous section. The quantitative agreement between simulations (black line) and the analytical result (red line, $\beta^* = 0.49 (k_B T)^{-1}$  and $\gamma^* = 1.80 (k_B T)^{-1}$) is not very good (Fig.~\ref{fig5}). 

This suggests that our assumption in the analytical model that the pool $\mathcal{T}$ of selected TCRs is uncorrelated, might not be good for short sequences. To test the effect of correlations, we generated a set of $K=10^3$ uncorrelated TCRs with amino acid frequencies $f_a^{(\rm{sel})}$ obtained from Eq.~(\ref{eq9}) and with MHC binding strengths drawn from the distribution in Eq.~(\ref{eq10}) and then checked the amino acid composition of foreign peptides recognized by this set. Figure~\ref{fig5} shows better agreement between the analytical result (red line) and simulations (blue data points) with uncorrelated TCRs. However, the analytical results still vary significantly from simulations. The large discrepancy is likely due to the inaccurate approximation of micro-canonical with the canonical ensemble of recognized foreign peptides for short peptides ($N=5$), which only holds in the limit of large peptides ($N \rightarrow \infty$). Worse agreement between the numerical (blue line) and analytical results (red line) for the amino acid composition of recognized peptides (Fig.~\ref{fig5}) compared to the results for the amino acid composition of selected TCRs (Fig.~\ref{fig2}b) might be due to lower magnitude of numerically obtained parameters $|\beta^*|=0.49 (k_B T)^{-1} < |\beta| = 3.06 (k_B T)^{-1}$ and the exponential dependence of amino acid frequencies on parameters $\beta$ and $\beta^*$ (Eqs.~(\ref{eq9}) and (\ref{eq15})). On the other hand the large difference between the numerically obtained results in black and blue lines is only due to the correlations of the selected TCRs in the thymus.

To examine this further, we have also tested the effect of correlations in the limit of a large number $K$ of selected T cells, since a mouse has  $\sim 10^8$ distinct T cells and a human has $\sim 10^9$ distinct T cells. Both numbers are larger than the total number of possible sequences ($20^N$ for $N=5$) in TCR peptide binding regions. This suggests that the sequence length $N$ should be larger ($N=6$ or $N=7$) or that TCRs differ also in regions that do not bind peptide, e.g. different TCR binding strengths to the same MHC type, different sets of TCR pools that correspond to different MHC types (each human has up to 12 different MHC types). For the values of $K$ that are of the order of the total number of sequences, the EVD $\Pi^*$  probes the tails of distribution $\rho^*\left( x | \vec s\, \right)$, where it is no longer Gaussian. Because the distribution $\rho^*\left( x | \vec s\, \right)$ is bounded, in the large $K$ limit the EVD approaches a delta-function centered at the $K$-independent value corresponding to the optimal binding energy:
\begin{equation}
E_0^*\left( \vec s\, \right) = \min_{\left\{ E_c, \vec t\, \right\} \in \mathcal{T}} \left \{ E_c + \sum_{i=1}^N J(t_i, s_i) \right\}.
\label{eq16}
\end{equation}
In the limit $N \rightarrow \infty$, a pool of TCR sequences is uncorrelated and the optimal binding energy can be written as:
\begin{equation}
E_0^* \left( \vec s\, \right) = E_{c,\rm{min}} + \sum_{i=1}^N J_{\rm{min}} (s_i),
\label{eq17}
\end{equation}
where $J_{\rm{min}} (a) = \min_{b} J(b,a)$. The condition for recognition of foreign peptides is
\begin{equation}
E_0^*\left( \vec s\, \right) < E_n,
\label{eq18}
\end{equation}
and
\begin{figure}[b]
\includegraphics[scale=.4]{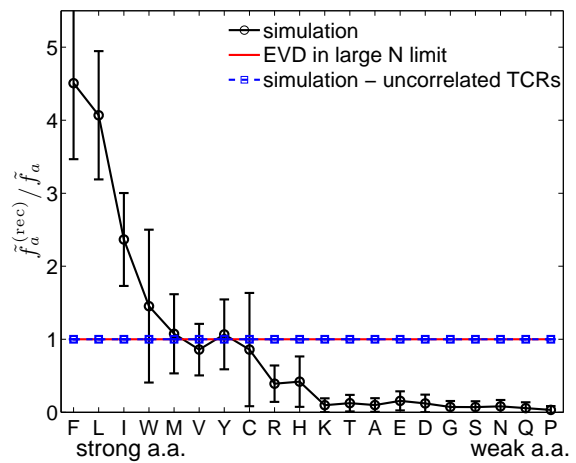}
\caption{
Amino-acid composition of foreign peptides recognized by a complete pool of selected TCRs,  The amino acids are ordered in decreasing frequency along the abscissa. The data points in black are obtained numerically where a complete pool of selected TCRs (obtained from thymic selection of all $20^N$ TCR sequences against $M=10^3$ self peptides) is challenged with \emph{L. monocytogenes} peptides (see text). The error bars reflect the differences for different realizations of $M=10^3$ self peptides. The red line is the result of the EVD analysis in the large $N$ limit from Eq.~(\ref{eq19}). The data points in blue are obtained numerically where a complete pool of uncorrelated TCRs, which is equal to the complete pool of $20^N$ sequences (see text), are challenged with \emph{L. monocytogenes} peptides. In this case, there are no error bars, because we use the complete pools of TCRs and foreign peptides with appropriate weights. In all cases we have used the Miyazawa-Jernigan matrix,  $J$~\cite{miyazawa96}, amino acid frequencies $f_a$ from humans (using the mouse proteome does not change the qualitative results~\cite{kosmrlj08}), and amino acid frequencies $\tilde{f}_a$ from the \emph{L. monocytogenes}~\cite{kosmrlj08, flicek08}.} 
\label{fig6}
\end{figure}
the probability for a sequence $\vec s$ to be recognized is governed by the Boltzmann weight $p\left( \vec s\, \right) \propto \left( \prod_{i=1}^N \tilde{f}_{s_i} \right) \exp\left[ - \beta^* E_0^* \left( \vec s\, \right) \right]$, where $\left\{ \tilde{f}_a \right\}$ are natural frequencies of different amino acids in the pathogen proteome, while the effect of TCR recognition is captured in the parameter $\beta^*$. The value of $\beta^*$ is determined by constraining the average energy $\langle E_0^* \left( \vec s \, \right) \rangle < E_n$, while maximizing entropy in the same manner as described before (see the paragraph after Eq.~(\ref{eq14})). The condition for $\beta^* > 0$ can be simplified to:
\begin{equation}
\frac{E_n - E_{c,\rm{min}}}{N} = \langle J_{\rm {min}} (a) \rangle = \frac{\sum_{a=1}^{20} J_{\rm {min}} \tilde{f}_a \exp\left[ - \beta^* J_{\rm {min}} (a) \right]}{\sum_{a=1}^{20} \tilde{f}_a \exp\left[ - \beta^* J_{\rm {min}} (a) \right]}.
\label{eq19}
\end{equation}
With the parameter $\beta^*$ determined, we can calculate the amino acid frequencies of recognized foreign peptides
\begin{equation}
\tilde{f}_a^{(\rm{rec})} = \frac{\tilde{f}_a \exp\left[ - \beta^* J_{\rm {min}} (a) \right]}{\sum_{b=1}^{20} \tilde{f}_b \exp\left[ - \beta^* J_{\rm {min}} (b) \right]}.
\label{eq20}
\end{equation}
For the relevant parameters in mouse (i.e. $N=5$, $E_p - E_n = 2.5 k_B T$, $E_{c,\rm{max}} = E_p - N \overline {J}$, $E_{c,\rm{min}} = E_n - N \overline{J}$, and $M=10^3$), we obtain $\beta^*=0$, which means that every foreign peptide is recognized. We tested this numerically by checking the properties of foreign peptides recognized by a complete pool of selected TCR. For each of the $20^N$ possible TCR sequences $\vec t$, we calculated the strongest interaction with $M=10^3$ self peptides $E_{\rm{min}} \left( \vec t\, \right) = \min_{\vec s \in M} \left\{ \sum_{i=1}^N J(t_i, s_i) \right \}$ and 
constructed an interval of $E_c$ values that could result in the TCR selection as $\left( E_{c,\rm{min}}^* , E_{c,\rm{max}}^* \right) = \left( E_n - E_{\rm{min}} \left( \vec t\, \right) , E_p - E_{\rm{min}} \left( \vec t \, \right) \right)$. The actually selected $E_c$ values for a given TCR sequence $\vec t$ are obtained by intersecting the $\left( E_{c,\rm{min}}^* , E_{c,\rm{max}}^* \right)$ interval with the allowed interval $\left( E_{c,\rm{min}} , E_{c,\rm{max}} \right)$ before selection. Thus we obtained a complete pool of selected TCRs with weights $\rho\left( E_c \right) \left( \prod_{i=1}^N f_{t_i} \right) \times \chi_{E_c \in (E_{c,{\rm min}} , E_{c, \rm{max}}) \cap (E_{c,{\rm min}}^* , E_{c, \rm{max}}^*)}$, where $\chi$ is an indicator function with a value of 1 if a TCR with given $E_c$ and $\vec t\, $  is selected and 0 otherwise. The complete pool of selected TCRs was then challenged against $10^5$ randomly generated foreign peptides that were representative of \emph{L. monocytogenes} and the whole process was repeated $1,000$ times with different realizations of $M=10^3$ self peptides. Fig.~\ref{fig6} shows a large disagreement between numerical simulations (black data points), where only foreign peptides with strongly interacting amino acids are recognized, and the analytical result (red line, Eq.~(\ref{eq20})), where every foreign peptide is recognized. Because the selected TCR sequences are correlated, we also tested the effect of correlations. Since $f_a^{(\rm{sel})}>0$ for every 
amino acid $a$ and $\rho^{(\rm{sel})} \left(E_c \right) > 0$ for all allowed $E_c$ values, the complete pool of uncorrelated TCRs correspond to all $20^N$ sequences and all possible interaction values with MHCs, $E_c$, with weights $\rho^{(\rm{sel})} \left ( E_c \right) \prod_{i=1}^N f_{t_i}^{(\rm{sel})}$. When a foreign peptide is tested against the complete pool of TCRs, it finds a TCR that results in optimal binding energy (Eq.~(\ref{eq17})). To numerically calculate the amino acid frequencies of foreign proteins recognized by complete uncorrelated pool of TCRs, we generate all possible $20^N$ foreign peptide sequences with appropriate weights $\left( \prod_{i=1}^N \tilde{f}_{s_i} \right)$. A foreign peptide sequence $\vec s\, $ is then recognized if the optimal binding energy (Eq.~(\ref{eq17})) exceeds the recognition threshold (Eq.~(\ref{eq18})). The numerical result for uncorrelated TCRs (blue data points) agrees very well with the analytical results (red line), which indicates that the correlations in selected TCR sequences (black data point) have an important role.

\subsection{Escape of autoimmune T cells}

\begin{figure*}[t]
\includegraphics[scale=0.5]{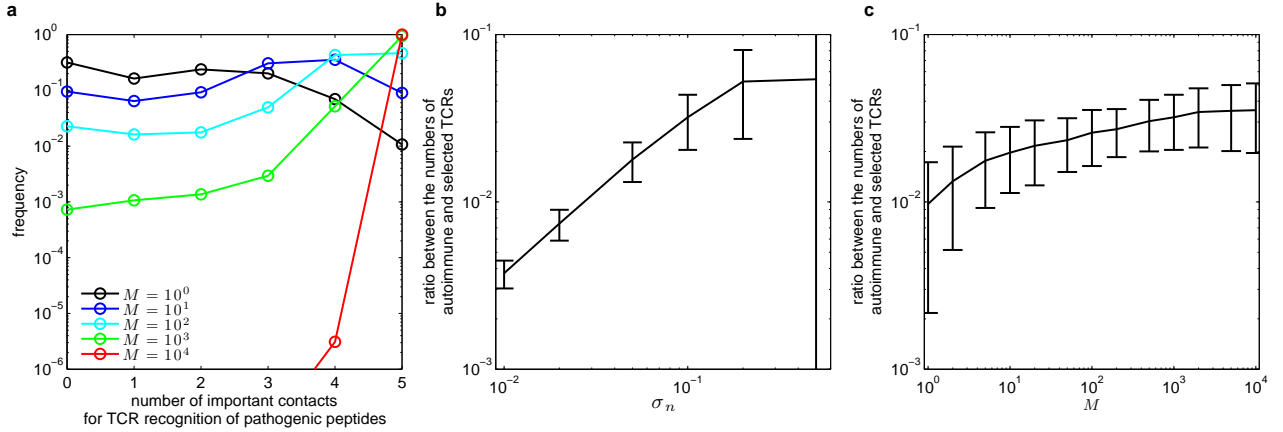}
\caption{
Thymic selection with soft thresholds for positive and negative selection. a) Histogram of the number of important contacts (defined in text) with which T cells recognize pathogenic peptides. This result and other characteristics of selected TCRs (data not shown) are qualitatively equivalent to the results obtained with sharp thresholds for positive and negative selection (Figs.~\ref{fig1} and \ref{fig3}). b) Ratio of the numbers of escaped autoimmune TCRs and selected TCRs as a function of the softness of the threshold for negative selection $\sigma_n$. Fraction of autoimmune TCRs increases with $\sigma_n$, until the softness of the thresholds becomes of the same order as the separation between the thresholds of positive negative selection, $\sigma_n + \sigma_p \sim E_p - E_n$. $\sigma_p = 1 k_B T$, $M=10^3$. c) Ratio of the numbers of escaped autoimmune TCRs and selected TCRs as a function of the number of types of self peptides ($M$). Fraction of autoimmune TCRs increases with $M$ and is roughly constant for large $M$, however the absolute numbers of TCRs are decreasing with $M$. $\sigma_p = 1k_B T$, $\sigma_n = 0.1 k_B T$. The error bars in (b) and (c) correspond to the standard deviation of the fractions of escaped TCRs obtained from repeating the thymic selection process many times.} 
\label{fig7}
\end{figure*}

Thymic selection is not perfect and autoimmune T cells, which interact strongly with self pMHCs, can escape from the thymus. Due to stochastic effects, it may happen that a diffusing T cell in the thymus never interacts with some peptides that would lead to negative selection. Also, even if a TCR binds strongly to a self-pMHC, it can escape with some probability because the negative selection threshold is not sharp. Here we only focus on the latter effect, which can be modeled with a soft threshold for negative selection. For a TCR $\vec t\,$ that interacts with self-peptide $\vec s\,$, the probability of negative selection is assumed to be
\begin{equation}
P_n\left( \vec t, \vec s\, \right) = \frac{1}{1 + \exp\left[ - \left( E_{\rm{int}} \left( \vec t, \vec s\, \right) - E_n \right) / \sigma_n \right]},
\label{eq21}
\end{equation}
where the parameter $\sigma_n$ denotes the softness of negative selection threshold. For a TCR that interacts strongly with self-pMHC ($E_{\rm{int}} < E_n$) the probability of negative selection is close to 1, while for a TCR that interacts with self-pMHC weakly ($E_{\rm{int}} > E_n$) the probability of negative selection is small. Similarly we can define the probability for positive selection $P_p$ with the corresponding softness $\sigma_p$. From experiments~\cite{daniels06} we know that the threshold for positive selection is softer ($\sigma_p > \sigma_n$). In this case thymic selection is modeled by testing each TCR sequence with all self-pMHCs: for each self-peptide we calculate the corresponding probabilities $P_p$ and $P_n$ of positive and negative selection, then we draw two uniformly distributed random numbers $r_p$ and $r_n$ from the $(0,1)$ interval and a TCR is positively (negatively) selected if $r_p < P_p$ ($r_n < P_n$). After thymic selection is completed, we check if any of the selected TCRs interacts strongly with any self-peptide ($E_{\rm{int}} < E_r = E_n$). Deterministic criteria are now used because strong interaction free energy leads to high probability of T cell activation. Any other deterministic or stochastic criteria would not qualitatively change the results. We find that the introduction of soft thresholds for positive and negative selection does not qualitatively change the results reported earlier regarding the composition of selected TCRs, etc. (Fig.~\ref{fig7}a and data not shown). Fig.~\ref{fig7}b shows that increasing the softness of the threshold for negative selection $\sigma_n$ increases the chance of escape of autoimmune TCRs. This is because strongly interacting TCRs are negatively selected with lower probability when the threshold for negative selection is softer. Interestingly, the ratio of the number of autoimmune T cells to the number of selected T cells seems to be roughly constant with the number $M$ of self-peptides used during the development in thymus (Fig.~\ref{fig7}c). The fraction of autoimmune T cells increases with $M$, but the rate of increase is small for large $M$. Note that with increasing number of self peptides $M$ both the nominator and denominator are decreasing, but the ratio is roughly constant.

\section{Discussion}

In this paper we extend our understanding of the problem of how the thymus designs a T cell repertoire that is both specific and degenerate for pathogenic peptide recognition. Previously~\cite{kosmrlj08, kosmrlj09, kosmrlj10} we argued that for selection against many self-peptides, negative selection imposes a constraint which results in selected TCR sequences composed of predominantly weakly interacting amino acids. We now find additionally that negative selection also results in selected TCRs that bind relatively weakly to MHC. But, interactions with MHC cannot be arbitrarily weak as that would prevent proper TCR orientation on MHC and peptide recognition.

Binding of such TCRs to pathogenic peptides is sufficiently strong for recognition only when pathogenic peptides are composed of predominantly strongly interacting amino acids. This is not too restrictive for the immune system, because several pathogenic peptides are derived from pathogenic proteins and presented to T cells. It is enough for T cells to recognize just a few pathogenic peptides, to activate the immune system and clear the infection. This may contribute to why there are only a few immunodominant peptides corresponding to any infection.

Equations~(\ref{eq15}) and (\ref{eq20}) provide an analytical expression that captures the characteristics of amino acids of the recognized foreign peptides. Figures~\ref{fig5} and \ref{fig6} show that the analytical result is not very accurate for short ($N = 5$) peptide sequences and we showed (Figs.~\ref{fig5} and \ref{fig6}) that this is due to the correlations in selected TCR sequences. In the future it would be interesting to study how (or if) these correlations vanish as peptide sequence length ($N$) is increased.

It is known that people who express a particular MHC type called HLA-B57 are more likely to control HIV infection than people without this MHC~\cite{deeks07, migueles00}. In a previous study~\cite{kosmrlj10} we found that HLA-B57 bind $\sim 6$ times fewer peptides than MHC molecules that are associated with faster progression to AIDS. This means that TCRs restricted for HLA-B57 are selected against fewer types of self peptides in the thymus, which results in a more cross-reactive T cell repertoire (Fig.~\ref{fig3}a and Fig.~\ref{fig7}a). In that study we showed that more cross-reactive T cell repertoire could contribute to better control of HIV infection. Interestingly, people expressing HLA-B57 are also more prone to autoimmune diseases~\cite{bhalerao98, chessman08}. We also studied the escape of autoimmune T cells from the thymus as a function of the number of self peptides $M$. While the ratio of escaped autoimmune T cells to selected T cells is roughly independent of $M$ in the relevant regime ($M\sim$  a few thousands, Fig.~\ref{fig7}c), the absolute numbers of escaped autoimmune T cells is higher. This implies that the rate of escape of autoimmune T cells could be higher in people expressing HLA-B57.

Note, however that allowing escape of autoimmune TCRs by having a soft threshold for negative selection does not alter our qualitative results regarding the characteristics of selected TCRs, and the origins of specific and degenerate TCR recognition of pathogen.

\acknowledgements{This work was supported by the Ragon Institute (A.K.C., A.K.), National Institutes of Health (NIH) Grant No. 1-PO1-AI071195-01 (A.K.C., M.K.), NSF Grant No. DMR-08-03315 (M.K.), and a NIH DirectorÕs Pioneer award (to A.K.C.).}

\bibliography{library}
\end{document}